# Dynamic reshaping of functional brain networks during visual object recognition

Rizkallah J.[1, 2], Benquet P.[1], Kabbara A. [1, 2], Dufor O. [4], Wendling F.[1], Hassan M.[1]

[1] Univ Rennes, LTSI, F-35000 Rennes, France

[2]AZM center-EDST, Lebanese University, Tripoli, Lebanon

[3] IMT Atlantique Bretagne Pays de la Loire, UMR CNRS Lab-STICC, Brest, France

Corresponding author:

Mahmoud Hassan

mahmoud.hassan@univ-rennes1.fr

# Abstract

Emerging evidence shows that the modular organization of the human brain allows for better and efficient cognitive performance. Many of these cognitive functions are very fast and occur in sub-second time scale such as the visual object recognition. Here, we investigate brain network modularity while controlling stimuli meaningfulness and measuring participant's reaction time. We particularly raised two questions: i) does the dynamic brain network modularity change during the recognition of meaningful and meaningless visual images? And ii) is there a correlation between network modularity and the reaction time of participants?

To tackle these issues, we collected dense-electroencephalography (EEG, 256 channels) data from 20 healthy human subjects performing a cognitive task consisting of naming meaningful (tools, animals…) and meaningless (scrambled) images. Functional brain networks in both categories were estimated at sub-second time scale using the EEG source connectivity method. By using multislice modularity algorithms, we tracked the reconfiguration of functional networks during the recognition of both meaningful and meaningless images. Results showed a difference in the module's characteristics of both conditions in term of integration (interactions between modules) and occurrence (probability on average of any two brain regions to fall in the same module during the task). Integration and occurrence were greater for meaningless than for meaningful images. Our findings revealed also that the occurrence within the right frontal regions and the left occipito-temporal can help to predict the ability of the brain to rapidly recognize and name visual stimuli. We speculate that these observations are applicable not only to other fast cognitive functions but also to detect fast disconnections that can occur in some brain disorders.

# Introduction

Information is continuously processed and integrated in the human brain. To ensure efficient cognitive function, complex brain networks are characterized by two key features. On the one hand, recent studies show that their modularity facilitates information processing as compared to non-modular organization (Sporns and Betzel 2016). Typically, in a large study of 77 cognitive tasks, Bertolero et al. showed that the presence of network modules (defined as a set of brain regions strongly connected to each other and weakly connected to the rest of the network) is correlated with different cognitive functions such as memory, visual processing and motor programming (Bertolero, Yeo et al. 2015).

On the other hand, emerging evidence shows that the dynamic behavior of brain networks is fundamental to understand cognition (Basset 2015). Typically, during a learning task, Basset et al. showed that the flexibility (defined as how often a given node changes its modular affiliation over time) of the networks facilitates the prediction of individual future performances in next learning sessions (Bassett, Wymbs et al. 2011). Both flexibility and integration (defined as the interactions between modules) in the frontal lobe during a 2-back working memory task were shown to be correlated with the performance accuracy (Braun, Schäfer et al. 2015).

All the above-mentioned studies were performed using fMRI, the spatial resolution of which allows for appropriate identification of brain areas involved in considered cognitive processing. However, most of cognitive processes occur on a very short duration and are likely to involve modular dynamic changes occurring at sub-second time scale. Unfortunately, such changes cannot be tracked with fMRI due to intrinsic time resolution (on the order of 1 second).

Therefore, our knowledge about the dynamic modifications of the modular organization of brain networks at a sub-second time-scale during cognitive activity remains elusive.

To tackle this issue, i.e. to assess how functional brain network modules dynamically reconfigure to ensure information processing and integration, we chose the well-defined visual object recognition and naming task (DiCarlo, Zoccolan et al. 2012) which involves fast cognitive processes (a few hundred of ms from stimulus onset to reaction). In order to guarantee sufficiently fast tracking of functional brain networks, we collected dense-electroencephalography (EEG, 256 channels) data from 20 healthy human subjects performing a cognitive task consist of naming meaningful (tools, animals…) and meaningless (scrambled) visual stimuli. Functional brain networks in both categories were estimated using 'dense-EEG source connectivity' method (Hassan and Wendling 2018). By applying multislice modularity algorithms to neocortical networks, we tracked the reconfiguration of brain modules at sub-second time scale during the recognition of these two object categories. Results showed that two relevant parameters, namely the integration and occurrence (defined as the probability on average of any two brain regions to fall in the same module during the task) parameters, exhibited stronger values for meaningless as compared with meaningful images. Our findings also revealed that the occurrence within the right frontal regions and the left occipito-temporal can help to predict the ability of the brain to rapidly recognize and name meaningful visual stimuli.

# Materials and methods

**Participants**

Twenty healthy volunteers (10 women and 10 men; mean age 23 y) with no neurological diseases participated in this study (the data of two females and 1 male participants were eliminated as EEG signals were very noisy due to electrodes impedance issues). 80 meaningful and 40 meaningless pictures (figure S1), taken from the Alario and Ferrand database (Alario and Ferrand 1999), were displayed on a screen as black drawings on a white background and the participants were asked to name the presented images. The same number of stimuli was used in the further analysis by selecting 40 meaningful images (the same for all subjects). E-Prime 2.0 software (Psychology Software Tools, Pittsburgh, PA) was used to display the pictures.

A typical trial started with a fixation cross that lasted 1sec, then the image was shown during 2.5s and followed by a blank screen for 1 or 2s (randomly selected) (Figure 1A, see also table S1 for more details about the images used in the study). The time between the picture onset and the beginning of vocalization recorded by the system was considered as naming latencies. The voice onset times were then analyzed using Praat software (Boersma 2001). The fastest response over trials and subjects was 535ms. Therefore, the analysis was performed from the stimulus onset up to 500ms following the stimulus in order to avoid muscle artifacts (due to articulation). Errors in naming were discarded for the following analysis. Subjects were informed about meaningless objects presence in the experiment and were instructed to say nothing when viewing them. All participants provided a written informed consent to participate in this study which was approved by the National Ethics Committee for the Protection of Persons (CPP), Braingraph study, agreement number (2014-A01461-46), and promoter: Rennes University Hospital.

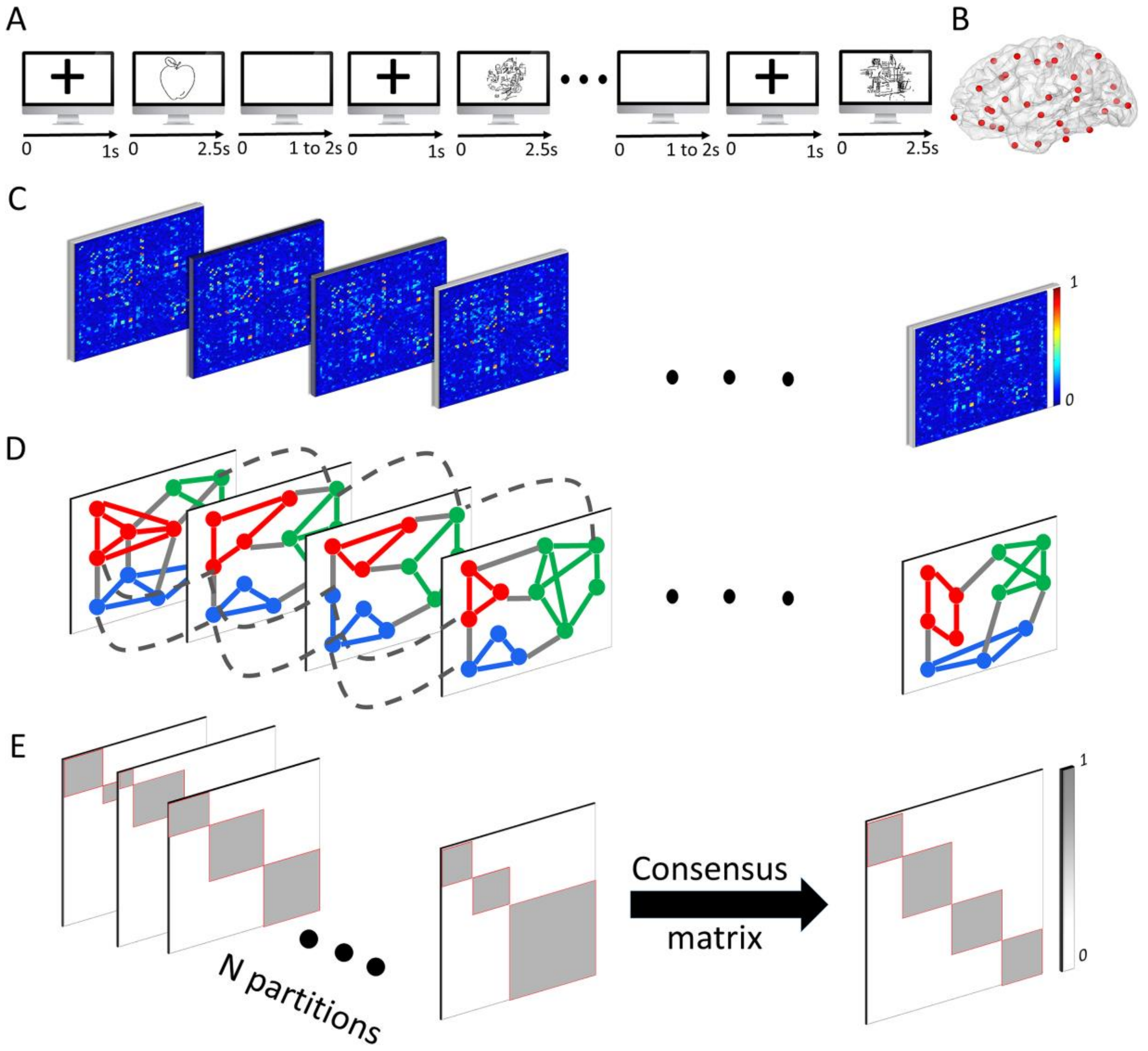

**Figure 1. Experimental setup. A. The cognitive task consisted in naming visual stimuli of two categories of: meaningful and meaningless pictures. B. 68 cortical regions defined from the Desikan-Killiany atlas (Desikan, Ségonne et al. 2006) were used. C. Functional connectivity was computed between these 68 regional time series for a period of 500 ms using the dense EEG source connectivity method in the gamma frequency band (30-45Hz). D. A dynamic module detection algorithm (Bassett, Wymbs et al. 2011, Bassett, Porter et al. 2013, Bassett, Yang et al. 2015) was used to identify network modules in each time window and to follow their evolution over time. Briefly, this algorithm constructs a modular allegiance matrix over the 100 runs at each window in which each element is 1 if two nodes are in the same module and is equal to zero otherwise. E. All matrices are then summed for each condition to obtain the global modular allegiance matrix, whose elements indicate the fraction of time windows in which nodes have been assigned to the same module. Then a community detection algorithm is applied to obtain a "consensus partition" as described in (Braun, Schäfer et al. 2015), which represents the common modular structure across all time windows.**

## Data recording and preprocessing

A dense EEG system (EGI, Electrical Geodesic Inc., 256 electrodes) was used to record brain activity. EEG signals were sampled at 1 KHz and then band-pass filtered between 0.3 and 45 Hz.

For the preprocessing, EEGLAB (Delorme and Makeig 2004) was used to reject and exclude the epochs contaminated by eye blinking (using ICA-based algorithm) and any other noise source. For some subjects, few electrodes with poor signal quality were identified. For these electrodes, the EEG signal was interpolated using signals recorded by surrounding electrodes.

In addition to dense EEGs, individual structural MRI was also available for each participant. A realistic head model was built by segmenting the anatomical MRI using Freesurfer (Fischl 2012). The individual MRI anatomy and EEGs were co-registered through identification of the same anatomical landmarks (left and right tragus and nasion). The lead field matrix was then computed for a cortical mesh of 15000 vertices using Brainstorm (Tadel, Baillet et al. 2011) and OpenMEEG (Gramfort, Papadopoulo et al. 2010). An atlas-based approach was used to project EEG signals onto a subject-specific anatomical framework consisting of 68 cortical regions (summarized in table S2 and visualized in figure 1B) identified by means of the Desikan-Killiany, (Desikan, Ségonne et al. 2006). Time series belonging to the same ROI were averaged after flipping the sign of sources with opposite directions.

**Construction of the functional networks**

Functional networks were constructed using the dense-EEG source connectivity method (Hassan, Dufor et al. 2014, Hassan, Benquet et al. 2015). This method includes two main steps: i) reconstruction of the temporal dynamics of the cortical regions from the scalp EEG signals and ii) measurement of the functional connectivity between reconstructed regional time series. The weighted Minimum Norm Estimate (wMNE) was used to reconstruct the cortical sources. A phase locking value (PLV) (Lachaux, Rodriguez et al. 1999) algorithm was then used to estimate the functional connectivity. Two versions of PLV were proposed by (Lachaux et al., 1999). The first method consists of looking at the "inter-trial" phase synchrony, which provides pair-wise connectivity values at each time point. This version works only in the case of task-related paradigms and in the presence of a high number of trials. The second version consists of choosing a sliding window, adapted to the analyzed frequency band. In the case of gamma band: 30-45 Hz, the duration of the smallest time window that contains a sufficient number of cycles (N=5) for PLV computation is ~0.133s. As shown by (Lachaux et al., 1999), both versions provide very similar results. In this study, we used the first version.

This choice of wMNE/PLV was supported by two comparative analyses performed in (Hassan, Dufor et al. 2014, Hassan, Merlet et al. 2017) that reported the superiority of wMNE/PLV over other combinations of five inverse algorithms and five connectivity measures. Briefly, in (Hassan, Merlet et al. 2017), the network identified by each of the inverse/connectivity combination used to identify cortical brain networks from scalp EEG was compared to a simulated network (ground truth). The combination that showed the highest similarity between scalp-EEG-based network and reference network (using a network similarity algorithm) was considered as the optimal combination. This was the case for the wMNE/PLV. The dense-EEG source connectivity method benefits from the intrinsically-excellent time resolution of the EEG. This lead to time-varying functional networks which spatio-temporal dynamics directly characterize the cognitive processes involved in considered task. Regarding technical details, readers can refer to (Kabbara, Falou et al. 2017, Kabbara, Eid et al. 2018) for detailed methodological description of the dense EEG source connectivity method as computed in this paper. The inverse solutions were computed using Brainstorm (Tadel, Baillet et al. 2011).

The wMNE/PLV combination was computed in the gamma band (30-45Hz). For each subject, this procedure yielded a set of weighted adjacency matrices describing the functional connectivity in each time window (Figure 1C).

Finally, not all elements of the connectivity matrix reflect significant functional relationships and that a threshold should be applied to retain only the ‘true’ functional connections. Here, we adopted the approach proposed in (Bassett, Wymbs et al. 2011). Briefly, for each connectivity matrix $A_{i,j}$, a *p*-value matrix $P_{i,j}$ was computed , based on the *t*-statistic**.** The computed *p*-values were corrected for multiple comparisons using the False Discovery Rate (FDR) approach of $p <$ 0.05. All $A_{i,j}$ whose *p*-values $P_{i,j}$ passed the statistical FDR threshold were retained (their values remained unchanged). Otherwise, the values were set to zero (Bassett, Wymbs et al. 2011) to build a thresholded weighted connectivity matrix.

**Multislice network modularity**

The resultant matrices were split into time-varying modules using a multislice community detection algorithm described in (Mucha, Richardson et al. 2010). This algorithm consists of introducing a coupling parameter that links nodes across slices (time windows) before performing the modularity maximization procedure (Figure 1D). This algorithm was recently applied on functional brain networks (Bassett, Wymbs et al. 2011, Bassett, Porter et al. 2013, Bassett, Yang et al. 2015). The multislice modularity is defined as:

$$Q_{ml} = \frac{1}{2\mu} \sum_{ijlr} \left\{ \left( A_{ijl} - \gamma_l \frac{k_{il} k_{jl}}{2m_l} \right) \delta_{lr} + \delta_{ij} C_{jlr} \right\} \delta\left( M_{il}, M_{jr} \right)$$

where nodes *i* and *j* are assigned to communities $M_{il}$ and $M_{jl}$ in layer *l*, respectively. $A_{ijl}$ represents the weight of the edge between these two nodes and $\gamma_l$ is the structural resolution parameter of layer *l*. $C_{jlr}$ is the connection strength between node *j* in layer *r* and node *j* in layer *l*. The structural resolution parameter $\gamma$ and the inter-layer coupling parameter *C* are usually set to 1. $k_{il}$ is the strength of node *i* in layer *l*, the $\delta$-function $\delta(x, y)$ is 1 if $x = y$ and 0 otherwise, $m = \frac{1}{2} \sum_{ij} A_{ij}$ and $\mu = \frac{1}{2}\sum_{jr} k_{jr}$ . This produces for every brain region at every time window a modular assignment reflecting the module allegiance.

The multilayer network modularity was computed 100 times as Q may vary from run to run, due to heuristics in the algorithm: each run can produce slightly different partitions of nodes into modules (Good, de Montjoye et al. 2010). To deal with this problem, we computed a consensus matrix, also called ‘co-classification matrix’ (Bassett, Porter et al. 2013, Fornito, Zalesky et al. 2016) whose elements indicate the ratio of each node to be in the same module with the other nodes among these 100 partitions (figure 1E). Only elements in the consensus matrix higher than an appropriate random null model were taken into account, as described in (Bassett, Wymbs et al. 2011). Finally, by applying the Louvain algorithm (Blondel, Guillaume et al. 2008) on the consensus matrix, we obtained a partition that is most representative of the network segregation. Brain regions repeatedly classified in the same module (over the runs) will have a high weight in the consensus matrix and will more likely be assigned to the same module after applying the modularity maximization algorithm. To investigate the consistency of the modules over time, we computed a final consensus matrix (same procedure as described above) by calculating the ratio of each node to be with the other nodes in the same module, among the time windows.

**Integration and occurrence metrics**

To quantitatively analyze the contribution of each module during the task, first we used the integration metric as described in (Bassett, Wymbs et al. 2011, Braun, Schäfer et al. 2015). Integration values reflect how modules are interacting with each other. It is computed as the average number of links each node in a given module has with the nodes in the other modules.

We calculated also a new metric called occurrence (%). In our case, the occurrence defined as the probability on average of any two nodes to fall in the same module over time, from 0 to 500ms.

This metric reflects the importance of the strong (or weak) temporal interactions between any two brain regions during the cognitive task.

**Software**

The connectivity measures, network measures and network visualization were performed using BCT (Rubinov and Sporns 2010), EEGNET (Hassan, Shamas et al. 2015) and BrainNet viewer (Xia, Wang et al. 2013), respectively. The Network Community Toolbox (http://commdetect.weebly.com/) was used to compute the consensus matrices as well as the values provided by the integration and flexibility metrics.

**Statistical test**

A Wilcoxon rank sum test was used to assess the statistical difference between the functional brain networks respectively associated to meaningful and meaningless objects. The difference between the two conditions was considered as significant when the *p*-value was less than 0.05. The Bonferroni method was used to correct for multiple comparisons.

# Results

### A. *Meaningful vs. meaningless networks*

For each participant, the network modularity was computed over time for both categories (meaningful and meaningless). A global consensus matrix was computed over all participants for each category, as described in the methods section. As depicted in figure 2A, the qualitative visual inspection of the obtained matrices, indicated that both conditions have a different modular configuration in term of network integration (connections outside the square red lines) and occurrence (weights of the matrix values).

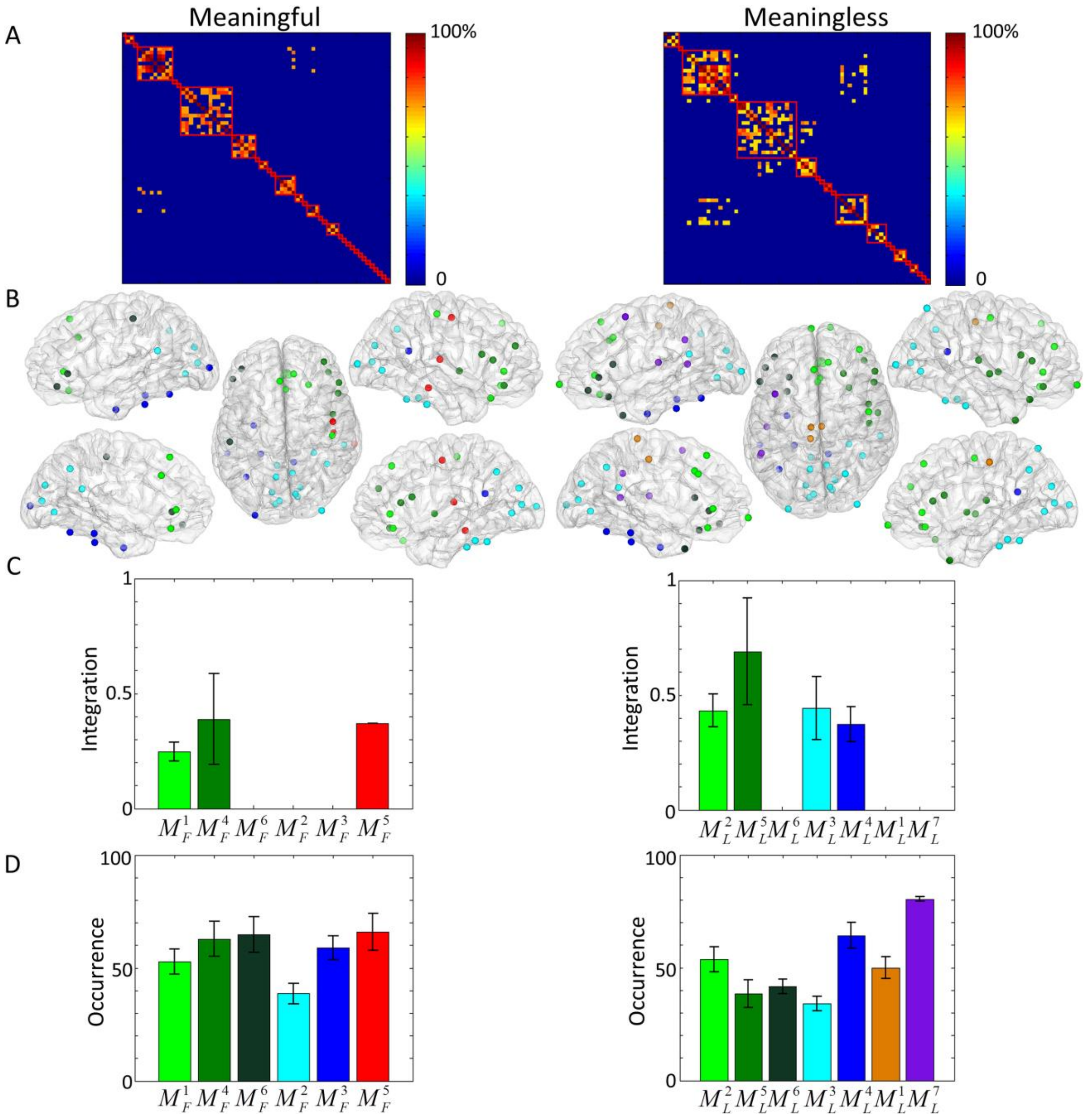

**Figure 2. A. The modular allegiance matrices for meaningful (left) and meaningless (right) conditions. B. A mapping of the different modules identified for both conditions (6 and 7 modules for the meaningful and meaningless categories). C. The values of integration for each module in both conditions. D. The values of occurrence for each module in both conditions. Note that modules having similar regions in both conditions are visualized by the same color. ($M_F$ and $M_L$ stand for meaningful and meaningless respectively).**

The resulting modules were projected on a 3D cortical surface, as illustrated in figure 2B. Modules having spatially-neighbor regions in both conditions are visualized by the same color and the region's names for each module are reported in table 1. To assess an average quantitative difference between the two categories, integration (figure 2C) and occurrence (figure 2D) values were computed from the final consensus matrix for each module. For both meaningful and meaningless objects, $\mathbf{M_F^4}$ and $\mathbf{M_L^5}$ modules located in the frontal and the temporo-frontal lobes respectively have the highest integration values. The temporo-central module $\mathbf{M_F^5}$ was interacting with other modules only for meaningful objects (figure 2C left) while the occipital and occipito-temporal modules ($\mathbf{M_L^3}$ and $\mathbf{M}_L^4$) were integrated only for meaningless objects (figure 2C right). Concerning the occurrence values, which represent the probability of any two brain regions to fall in the same module during the task, results showed fewer differences in both conditions (Figure 2D). Globally, the $\mathbf{M_F^1}$, $\mathbf{M_F^4}$ and $\mathbf{M_F^6}$ modules (located mainly in the frontal cortex) showed higher occurrence values for meaningful than meaningless frontal modules ($\mathbf{M_L^2}$, $\mathbf{M_L^5}$ and $\mathbf{M_L^6}$).

| Meaningful | | Meaningless | |
|---|---|---|---|
| $M_F^1$ | caudalanteriorcingulate.R (cACC.R), lateralorbitofrontal.R (LOF.R), medialorbitofrontal.L/R (MOF.L/R), postcentral.R (postC.R), rostralanteriorcingulate.L/R (rACC.L/R), superiorfrontal.L/R (sFG.L/R) | $M_L^2$ | caudalanteriorcingulate.L/R (cACC.L/R), frontalpole.L (FP.L), frontalpole.R (FP.R), medialorbitofrontal.L/R (MOF.L/R), precentral.R (preC.R), rostralanteriorcingulate.L/R (rACC.L/R), rostralmiddlefrontal (rMFG.R), superiorfrontal.L/R (sFG.L/R), |
| $M_F^2$ | cuneus.L/R (CUN.L/R), fusiform.R (FUS.R), inferiortemporal.R (ITG.R), isthmuscingulate.L (iCC.L), lateraloccipital.R (LOG.R), lingual.L/R (LING.L/R), parahippocampal.R (paraH.R), pericalcarine.L/R (periCAL.L/R), precuneus.L/R (PCUN.L/R) | $M_L^3$ | cuneus.L/R (CUN.L/R), fusiform.R (FUS.R), inferiortemporal.R (ITG.R), isthmuscingulate.L (iCC.L), lateraloccipital.L/R (LOG.L/R), lingual.L/R (LING.L/R), parahippocampal.R (paraH.R), pericalcarine.L/R (periCAL.L/R), precuneus.L/R (PCUN.L/R), superiorparietal.R (SPL.R) |
| $M_F^3$ | entorhinal.L (ENT.L), fusiform.L (FUS.L), inferiortemporal.L (ITG.L), isthmuscingulate.R (iCC.R), lateraloccipital.L (LOG.L), parahippocampal.L (paraH.L) | $M_L^4$ | entorhinal.L (ENT.L), fusiform.L (FUS.L), inferiortemporal.L (ITG.L), isthmuscingulate.R (iCC.R), parahippocampal.L (paraH.L) |
| $M_F^4$ | Insula.R (INS.R), parsopercularis.R (pOPER.R), parsorbitalis.R (pORB.R), parstriangularis.R (pTRI.R), rostralmiddlefrontal.R (rMFG.R) | $M_L^5$ | Insula.R (INS.R), lateralorbitofrontal.R (LOF.R), parsopercularis.R (pOPER.R), parsorbitalis.R (pORB.R), parstriangularis.R (pTRI.R), superiortemporal.R (STG.R), temporalpole.R (TP.R), transversetemporal.R (TT.R) |
| $M_F^5$ | middletemporal.R (MTG.R), precentral.R (preC.R), transversetemporal.R (TT.R) | $M_L^1$ | paracentral.L (paraC.L), posteriorcingulate.L/R (PCC.L/R) |
| $M_F^6$ | parsorbitalis.L (pORB.L), parstriangularis.L (pTRI.L), postcentral.L (postC.L) | $M_L^6$ | lateralorbitofrontal.L (LOF.L), parsopercularis.L (pOPER.L), parsorbitalis.L (pORB.L), parstriangularis.L (pTRI.L), temporalpole.L (TP.L) |
| | | $M_L^7$ | bankssts.L (BSTS.L), caudalmiddlefrontal.L (cMFG.L), supramarginal.L (SMAR.L), transversetemporal.L (TT.L) |

**Table 1: Modules obtained for both categories (meaningful and meaningless) in the consensus matrices.**

We then computed the statistical difference between the two conditions in term of integration and occurrence at the level of each brain region. Significant differences were obtained at three brain regions: the left isthmus cingulate (iCC.L), the right lingual (LING.R) and the right fusiform (FUS.R) as illustrated in Figure 3A ($p<0.01$, *uncorrected for multiple comparisons)*. Significant differences in the occurrence values were also obtained at four edges connecting the right caudal anterior cingulate (cACC.R) with the right pars opercularis (pOPER.R), the right cuneus (CUN.R) with the right pars opercularis (pOPER.R), the left lateral occipital (LOG.L) with the right pars opercularis (pOPER.R) and the right parahippocampal (paraH.R) with the right pars obitalis (pORB.R), as illustrated in Figure 3B ($p<0.01$, *uncorrected for multiple comparisons)*. Interestingly, these occipito-frontal and temporo-frontal connections were mainly located in the right hemisphere.

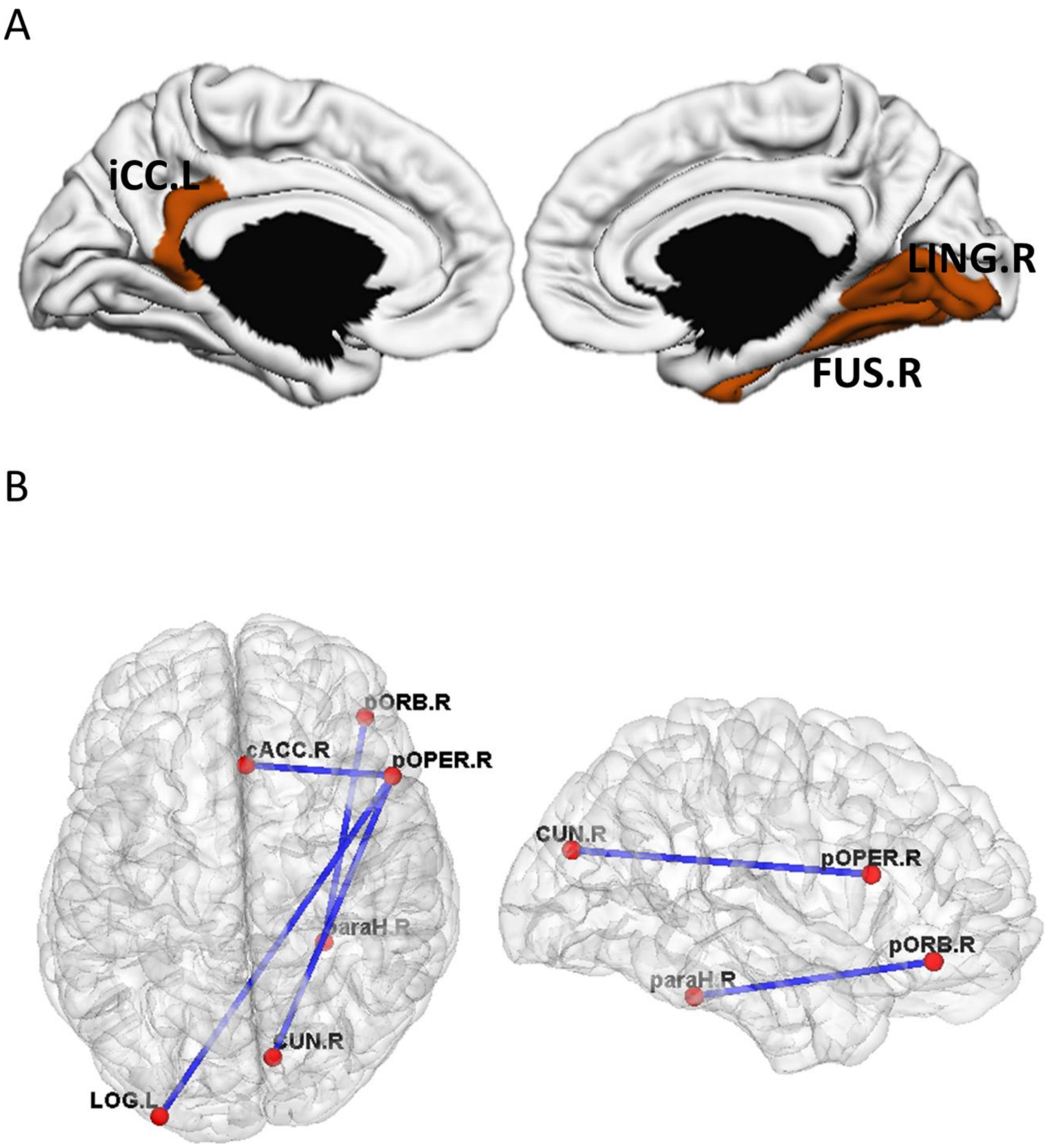

**Figure 3: A. Brain regions showing significant differences (meaningless > meaningful) in term of integration (p<0.01, uncorrected for multiple comparisons). B. Brain connections showing significant difference (meaningless > meaningful) in term of occurrence (*p*<0.01, uncorrected for multiple comparisons). See table S2 for the full name of the reported regions.**

### *B. Correlation between dynamic modularity and reaction time*

Here, we explored the correlation that may exist between the network modularity and the participant's reaction time, available only for 12 participants and defined as the time interval between the stimulus onset and the instant when participants start to name the displayed picture. We computed the values of the integration and the occurrence metrics for each region and edge, respectively. This part was realized only on the meaningful pictures as for meaningless pictures participants were asked to say nothing. Results (not shown here) revealed no significant correlation between the module's integration and the reaction times. However, the occurrence computed for each link showed positive and negative correlations at a set of connections. Figure 4A shows the connections where significant correlations ($p<0.01$, *uncorrected for multiple comparisons*) between the occurrence and the reaction time were observed. The positive correlations (green lines) were mainly located at the occipito-temporal regions while the negative correlations (red lines) were mainly located at the frontal cortex. By averaging the values of those edges (their occurrence), Figure 4B shows the strong correlations in both negative ($\rho=-0.9$, $p<0.001$) and positive ($\rho=-0.83$, $p=0.001$) cases.

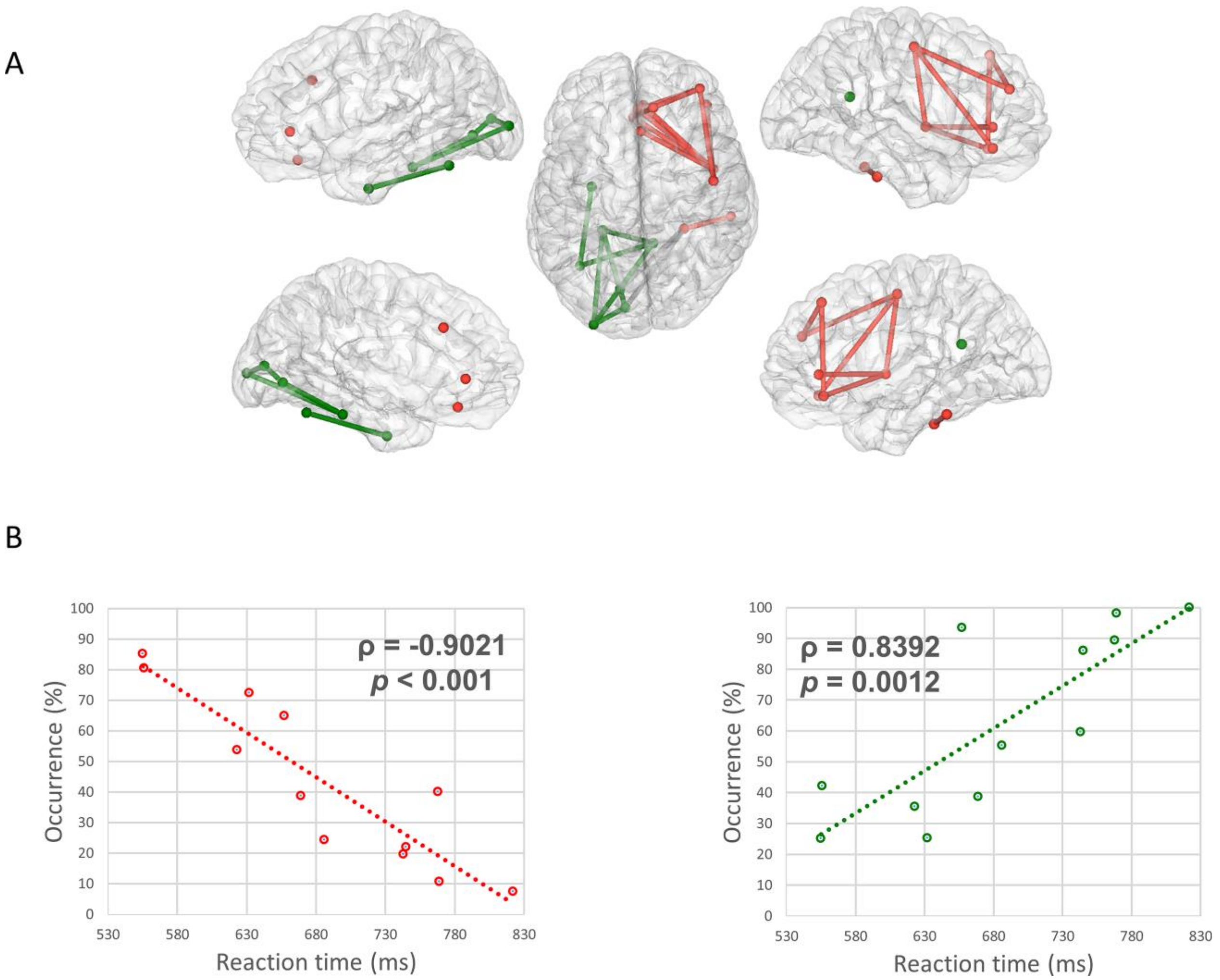

**Figure 4. A. Brain regions of 12 participants that showed significant correlation between their occurrence values and the participant's reaction time B. Correlation between the average occurrence over these connections and the minimal reaction time for each participant. (Note that this part was realized only on the meaningful pictures as for meaningless pictures participants were asked to say nothing and for only 12 participants as the reaction time was only available for them.)**

# Discussion

Emerging evidence shows that the human brain network is essentially organized into modules allowing the required adaptability to external drivers without necessarily modifying underlying structures (Braun, Schäfer et al. 2015). In a previous preliminary work (performed in a 'static' way), we showed that the network modularity could be a powerful tool to explore the overall reconfiguration of brain networks during visual object identification (Rizkallah, Benquet et al.

2016, Mheich, Hassan et al. 2017). Yet, this modular architecture is dynamic and may rapidly reshape according to external stimuli. Here, we showed that the functional connectivity during object recognition is dynamically organized into adapting modules and that this adapting modular architecture could be related to the ‘meaning’ of the visual stimuli. By quantifying these dynamic modules, we showed that network modularity (in term of occurrence) can predict the speed of object recognition and naming. Results are further discussed hereafter.

**Brain network modularity and cognitive functions**

The modular architecture of the human brain networks was reported using various neuroimaging techniques and at different scales (Sporns and Betzel 2016). It was firstly revealed on large scale structure network by Hagmann et al. (Hagmann, Cammoun et al. 2008). Regarding the functional role of brain network modules, numerous studies revealed that the brain is organized into a set of modules of functionally interconnected regions (Meunier, Lambiotte et al. 2010, Power, Cohen et al. 2011). The modular architecture of the human brain was shown to be related to individual cognitive state (Andric and Hasson 2015), development (Meunier, Achard et al. 2009) and brain diseases (Fornito, Zalesky et al. 2015). In addition, it was revealed that network modular architecture provides more flexible learning and promotes functional specialization (Bassett, Yang et al. 2015, Sporns and Betzel 2016).

Here, we investigated the possible functional role of brain network modularity during recognition (and naming) of meaningful and meaningless images. We showed that the two conditions have different modular organization mainly in the module’s node integration and module’s edge occurrence.

Our findings showed higher integration and occurrence values for meaningless compared to meaningful images. Higher integration for meaningless images was observed at right lingual and right fusiform regions. Interestingly, these region are known for their essential role in visual processing (Cai, Chong et al. 2015). For instance, in a PET study aimed at identifying brain structures involved in animal picture identification by comparison to meaningless shapes, results showed a primary activation of the lingual gyrus and the fusiform gyrus (Perani, Schnur et al. 1999). Our results also suggest that the left isthmus also shows significantly different integration between the two conditions (meaningless > meaningful). A possible explanation relates to the role of this region in the regulation of the focus of attention (Leech and Sharp 2014).

The increase in module integration for meaningless objects can be interpreted as an enhancement of communication between nodes, due to multiple attempts to match the detected image characteristics present in the visual stimuli with object representations already encoded in the memory involving the ventral visual pathway. Indeed, the complex shape information of scrambled images cannot be instantly linked to a recognized object in the brain database. Thus, searching for coherent signal (corresponding image) through visual pathway would increase module communication from the low/intermediate (image reconstruction) to high level of visual processing (image identification).

In term of occurrence, the occipito-frontal connections were found to be higher for meaningless than meaningful objects. As the occurrence is defined as the probability of two brain regions to stay in the same module during the task, thus it also provides a measure of the duration of communication between these two regions. A possible interpretation of higher occurrence values is that the coherent communication in networks involved in object identification (visual pathway) and decision making in case of unclear choice (prefrontal cortex) is not immediate for meaningless images. Indeed, the difficulty to identify and the unsuccessful multiple trials to match the image might increase the number of times the brain regions inter-communicate, and thus the occurrence values. Interestingly, denser scalp-EEG networks for meaningless images compared to meaningful images were also reported in (Gruber and Müller 2005). The lateralization of the connections during object recognition was also reported previously and several clinical studies linked the right hemisphere with object recognition. For instance, studies showed that patients with right-hemisphere disease (RHD) performed significantly worse than subjects with left-hemisphere disease (LHD) when asked to discriminate between faces, to discriminate between emotional faces, and to name emotional scenes (DeKosky, Heilman et al. 1980).

**Methodological issues**

First, we computed the functional networks at the EEG gamma band (30-45 Hz). This choice was supported by previous studies in the field of object recognition and naming. These studies reported that gamma is the most involved frequency band in the information processing during this cognitive task (Rodriguez, George et al. 1999, Hassan, Benquet et al. 2015, Liljeström, Kujala et al. 2015). Nevertheless, other frequency bands (and some frequency-frequency couplings), can be indeed involved in other aspects of this task such as the memory process. This issue is however beyond the scope of this study.

Second, the muscle artifact in the gamma band is a serious methodological issue. Here, we have reduced the effect of the muscle artifacts by i) asking the participants to not moving as much as possible ii) limiting our network-based dynamic analysis between the onset (presentation of the visual stimuli) to the moment before the fastest responses, in order to avoid the muscle artifacts due to articulation (naming) and iii) selecting the low gamma band (<45Hz) which is less contaminated by muscle artifacts comparing to high gamma (>50Hz).

Third, although significant results were obtained for region-wise analysis (integration) or edges-wise analysis (occurrence), these results did not resist the correction for the multiple comparisons. Thus, results should be interpreted with caution. Nevertheless, the results showed very strong correlations (> 0.8) values between the network metrics (mainly the occurrence) and the reaction time.

Fourth, a variety of thresholding methods are available, but none is free of bias. For example, one could apply a threshold that ensures only the top 10 % of connections across a sample of individuals are retained. In this case, it is recommended to repeat analyses across a range of threshold (values and approaches) to ensure that any results obtained are robust to this methodological parameter. Here we adopted the automatic threshold method used in Basset el al. (Bassett, Wymbs et al. 2011), where authors dealt also with the dynamics of functional brain networks. The main advantage of this method is that it is based on statistical tests and not an arbitrary choice of the threshold value.

Fifth, it is important to keep in mind that measuring the functional connectivity is generally corrupted by the volume conduction problem, a known problem regarding functional couplings at the scalp level (Schoffelen and Gross 2009). Therefore, connectivity analysis at source level was shown to reduce the effect of volume conduction as connectivity methods are applied to “local” time-series (analogous to local field potentials) generated by cortical neuronal assemblies modelled as current dipole sources. Nevertheless, these so-called “mixing effects” can also occur in the source space but can be reduced by an appropriate choice of connectivity measures. Indeed, false functional couplings can be generated by some connectivity methods when applied to mixed signals such as estimated brain sources. To address this issue, a number of methods were developed based on the rejection of zero-lag correlation. In particular, “unmixing” methods, called “leakage correction” (including the orthogonalization approach), have been reported which force the reconstructed signals to have zero cross-correlation at lag zero (Colclough, Brookes et al. 2015). Although handling this problem -theoretically- helps interpretation, a very recent study showed that the current correction methods also produce erroneous human connectomes under very broad conditions (Pascual-Marqui, Biscay et al. 2017). In addition, several experimental studies reported the presence of zero-lag corrections in the human brain. We believe that there is no ideal solution yet for this issue and that further methodological efforts are needed to completely solve the spatial leakage problem.

***Behavioral significance of network modularity***

Several functional network modularity-based metrics have previously been linked to many behavioral modifications in brain network dynamics in response to cognitive difficulties by predicting individual differences in learning (Bassett, Wymbs et al. 2011), working memory (Braun, Schäfer et al. 2015) and other cognitive tasks (Cole, Reynolds et al. 2013, Mattar, Betzel et al. 2016). Here we showed that the network occurrence, essentially for prefrontal brain regions, was negatively correlated with the reaction time (defined as the time from the stimulus onset to the instant when participant starts the naming process). On the other side, a positive correlation between the occurrence values of occipito-temporal connections was observed with the reaction time.

The occurrence values, that reflect the duration of communication between brain regions, increased in the occipito-temporal ventral visual pathway (including lingual cortex, lateral occipital cortex, enthorinal cortex, fusiform gyrus and parahippocampal cortex) which is known to be involved in the categorization and identification of the visual stimuli (Clarke and Tyler 2015), along with the reaction time. This result suggests that the longer it takes to identify the object, the longer is the occurrence in the occipito-temporal pathway and the longer is the reaction time.

Conversely, an immediate recognition and identification of a visual stimuli is associated with a decrease of occurrence values between brain regions involved in task-conflict (anterior cingulate (Shenhav, Straccia et al. 2014)) and decisions associated to image identification such as the frontal cortex (namely the superior frontal cortex, rostral middle frontal cortex, medial orbitofrontal cortex and precentral cortex) (Gilbert and Li 2013). This result suggests a decreased processing time of cognitive information as soon as the image is identified; the participant is ready to name the proper object and consequently a shorter reaction time. It is noteworthy that these results and interpretations should be taken with some cautions as the number of subjects is relatively small (N=12).

Finally, this prediction of the visual object recognition time can be easily extended not only to other cognitive tasks using other modalities (auditory, motor) but also to understand the possible modifications of network dynamics in patients with brain disorders, an issue that was recently discussed using other network modularity metrics (Mattar, Betzel et al. 2016, Zhang, Cheng et al. 2016).

# Acknowledgment

This work was supported by the Rennes University Hospital (COREC Project named BrainGraph, 2015-17). The work has received a French government support granted to the CominLabs excellence laboratory and managed by the National Research Agency in the "Investing for the Future" program under reference ANR-10-LABX-07-01. This work was also supported by the European Research Council under the European Union's Seventh Framework Programme (FP7/2007-2013) / ERC grant agreement n° 290901.